# Comparison of mechanical and molecular measures of mobility during constant strain rate deformation of a PMMA glass

Benjamin Bending[#] and M.D. Ediger*

Department of Chemistry, University of Wisconsin-Madison, Madison, WI 53706, USA

[#]Present address: Electronic Materials Solutions Division, 3M Center, St. Paul, MN 55144-1000 USA

*To whom correspondences should be addressed

(To be submitted to J. Polym. Sci. Part B: Polym. Phys. Ed.)

**Abstract**

We performed constant strain rate deformation and stress relaxation on a poly(methyl methacrylate) glass at $T_g$ – 19 K, utilizing three strain rates and initiating the stress relaxation over a large range of strain values. Following previous workers, we interpret the initial rate of decay of the stress during the relaxation experiment as a purely mechanical measure of mobility for the system. In our experiments, the mechanical mobility obtained in this manner changes by less than a factor of 3 prior to yield. During these mechanical experiments, we also performed an optical measurement of segmental mobility based upon the reorientation of a molecular probe; we observe that the probe mobility increases up to a factor of 100 prior to yield. In the post-yield regime, in contrast, the mobilities determined mechanically and by probe reorientation are quite similar and show a similar dependence upon the strain rate. Dynamic heterogeneity is found to initially decrease during constant strain rate deformation and then remain constant in the post-yield regime. These combined observations of mechanical mobility, probe mobility, and dynamic heterogeneity present a challenge for theoretical modeling of polymer glass deformation.

## Introduction

Deformation properties often dictate what polymer glass can be used in particular applications. In the limit of very small deformation, polymer glasses exhibit a linear response regime in which they respond as elastic solids. If a constant strain rate deformation is continued beyond this regime, the polymer glass may yield, undergo strain softening, and finally exhibit strain hardening if failure does not occur first.[1-3] As an example of the limits of our present understanding, it is difficult to predict the yield stress as this quantity exhibits a complex dependence on temperature, strain rate, and the thermal and mechanical history of the glass.[4-5] A major impediment to a full understanding of polymer glass deformation is the highly non-linear nature of the process. If a polymer glass responded as a linear viscoelastic material during a typical deformation, the flow stress would be more than 100 times larger than is observed.

An important indication of the non-linearity of a typical deformation experiment on a polymer glass is the change of the segmental relaxation time during deformation. By definition, in the linear response regime, molecular relaxation times are not altered by deformation. In contrast, there is strong evidence that segmental relaxation times during deformation can decrease by a factor of 100 or more.[6-11] In many theories and models of polymer glass deformation, this acceleration of segmental motion is the key reason for the nonlinear mechanical response.[1, 5, 12-20] Computer simulations are also consistent with this view.[21-24] As a crude approximation, one could say that stress induces enhanced segmental motion and this in turn allows the polymer glass to flow under conditions where it would otherwise act as an elastic solid. Given the central importance of segmental mobility in understanding the deformation of polymer glasses, it is not surprising that many efforts have been made to measure this mobility directly and indirectly during deformation.

A number of investigators have used mechanical stress relaxation measurements in order to infer changes in molecular mobility in polymer glasses during nonlinear deformation. For example, Yee et al. performed stress relaxation experiments on glassy polycarbonate at room temperature.[25] They observed that the stress relaxation time was much longer after a very small strain (~0.1%) than after strains of 1-5%. Qualitatively, these results could be interpreted to indicate that molecular motion was accelerated by deformation outside the linear response regime and that the time scale for stress relaxation was accelerated as a result. Yee et al. also performed more sophisticated experiments in which a small strain step (~0.1%) was added to a larger strain and the response to this smaller deformation was used to determine a characteristic relaxation time for the glass. This "tickle" experiment was interpreted to indicate that nonlinear deformation at constant temperature decreased molecular relaxation times in a manner that is similar to the effect of increasing the temperature in the absence of deformation. McKenna and Zapas performed similar experiments on a PMMA glass and obtained qualitatively similar results.[26] However, these later authors argued that the "tickle" experiment could not be rigorously interpreted as indicating enhanced molecular mobility and that such data should be interpreted in the context of a nonlinear deformation model. Caruthers and coworkers used constant strain rate deformation followed by stress relaxation on glasses of PMMA and an epoxy to infer changes in segmental mobility during deformation using the initial slope of the stress relaxation decay.[27-29] In this case, comparison was made with a model[14, 30] that predicts both enhanced molecular mobility and a change in the distribution of segmental relaxation times during deformation. Very recently, Liu et al. performed stress relaxation experiments after constant strain rate deformation in aged and mechanically rejuvenated glasses of polycarbonate, polystyrene, and poly(methyl methacrylate).[31] Remarkably, they reported that the stress relaxation curves exhibited a nearly universal decay after multiplying the observation time by the deformation rate used prior to stress relaxation; these results were also interpreted in terms of enhanced segmental mobility.

In light of the potential ambiguity in determining changes in molecular mobility from purely mechanical measurements, methods that more directly sense segmental motion have been utilized to gain an understanding of polymer deformation. One of the first molecular observations of increased mobility during deformation was seen in the enhanced diffusion of a diluent in a polymer glass.[7] Loo et al. used solid state NMR to observe enhanced molecular motion during the deformation of semi-crystalline nylon.[6] In a related observation, Watanabe utilized a dielectric technique to determine that segmental relaxation in a polymer melt shifts towards shorter times during shear flow.[32] Lee et al. have used the reorientation of a molecular probe to sense changes in segmental mobility of a polymer glass during constant stress deformation. [8-10, 33] These probe reorientation measurements were later extended to constant strain rate deformation.[11] Kalfus et al.[34] and Perez-Aparicio et al.[35] recently demonstrated dielectric measurements during tensile deformation. While all of these molecular measurements indicate enhanced segmental mobility during deformation, only the probe reorientation measurements have quantified how the average segmental relaxation time evolves during deformation.

Our goal in this work is to compare segmental mobility during the deformation of a polymer glass, as sensed by probe reorientation, with a mechanical relaxation time determined from a stress relaxation measurement. While the methods used to measure molecular mobility during deformation are specialized and have only been carried out in a few laboratories, mechanical measurements of mobility are accessible to a large number of laboratories. As mobility during deformation appears to be the central quantity needed to understand polymer glass deformation, it is of interest to understand under what conditions this quantity can be determined from a purely mechanical experiment.

Here we present constant strain rate deformation experiments on lightly crosslinked poly(methyl methacrylate) glasses, followed by stress relaxation. Following the protocol developed by Lee et al.[27] we determine a mechanical relaxation time from the initial decay of the stress relaxation. These experiments were performed at $T_g$ -19 K at three different strain rates with stress relaxation being initiated over a large range of strain values. We also perform optical measurements of probe reorientation throughout the constant strain rate deformation and stress relaxation, thus allowing a detailed comparison of the probe reorientation times with the mechanical relaxation times.

We find that the relationship between the mechanical and probe relaxation times differs in the pre-yield and post-yield regimes. In the post-yield regime, the two relaxation times are very similar (differing at most by a factor of ~2) and both exhibit similar power law relationships with the strain rate with exponents near -0.8. In the pre-yield regime, the probe and mechanical relaxation times differ qualitatively; as yield is approached, the mechanical relaxation times are almost constant while the probe relaxation times decrease by a factor of 100 or more. We interpret the results as indicating that the probe reorientation times accurately measure the changes in the average segmental relaxation time that occur during deformation. The mechanical relaxation time determined from the initial stress relaxation is more complex and changes as a result of additional factors that we discuss below. We also observe the evolution of the width of the relaxation time distribution during deformation and find a significant narrowing of the distribution in the pre-yield regime. The combined observations of mechanical relaxation times, probe relaxation times, and changes in the relaxation time distribution present a challenge for theoretical modeling of polymer glass deformation.

## Experimental Methods

Sample Preparation

Films of lightly crosslinked PMMA were prepared via bulk free radical polymerization using a procedure similar to that described previously.[11, 33] Ethylene glycol dimethacrylate (1.5 wgt %) was used as the crosslinking agent. The optical probe, N,N'-dipentyl-3,4,9,10-perylenedicarboximide (DPPC), is present at a concentration of ~ $10^{-6}$ M. Benzoyl peroxide at 0.1 wgt % was used to thermally initiate radical polymerization at 70°C for 30 minutes. The resulting viscous fluid was squeezed between microscope slides with aluminum foil spacers and then fully polymerized in a nitrogen purged oven, first at 70°C and then at 120°C. Films prepared in this manner had a $T_g$ of 392 ± 1 K (onset value from the second DSC heating scan at a rate of 10 K/min).

PMMA films were die cut into 'dog bone' style samples 33 mm in length and 2 mm in width. The film thickness was 20-30 μm in the center with the thickness increasing gradually to around 40-60 μm towards the ends of the sample. The optical experiment was performed near the thinnest portion of the sample, where yielding initiates. The sample was clamped into the deformation apparatus using sandpaper grips as described in a previous publication.[11]

Constant Strain Rate Deformation and Stress Relaxation

Prior to deformation, each PMMA sample was annealed at $T_g$+20 K for at least 2 hours with no stress applied. The sample was then cooled to the testing temperature of 373 K ($T_g$ – 19 K) at a rate of 1 K/min and annealed at this temperature for 30 minutes prior to deformation. This thermal history was used for all experiments in this paper. A single PMMA sample was used to obtain the data presented here but data acquired on other samples are consistent with our conclusions. The thermal annealing described above was adequate to erase memory of the previous deformation and to return the global and local strain to a zero strain value.

During tensile deformation we control the global strain rate by adjusting the velocity of a linear actuator. We measure the local strain by bleaching lines about 300 μm apart on the sample and optically tracking their separation as the sample is deformed. Figure 1 illustrates the behavior of the local strain for one experiment. For this experiment, we measured the local strain about six times more often than for a typical experiment in order to allow a detailed comparison between the global and local strain. As shown in the inset of figure 1, the local strain matches the global strain almost to yield (which occurred at 0.028 strain). In the post-yield regime of this experiment, the local strain rate is about two times larger than the global strain rate. We fit a $7^{th}$ order polynomial to the local strain (black line in figure 1) in order to obtain a continuous and smooth function of strain against time for later analysis. The average local strain rate in the post-yield region is determined from the derivative of this fit by averaging the strain rate from yield up to a local strain of 0.075. For all the experiments reported in this paper, the local strain rate in the post-yield regime was 2-3 times larger than the global strain rate. The measurements reported here do not extend into the strain hardening regime; a limited exploration of this regime is presented in reference [11].

The precise relationship between global and local strain rate is controlled by the position on the sample where the local strain is measured. For the stress relaxation experiments presented in this paper, the measurements of the local strain (and probe mobility) were performed at nearly the same location on the sample for all experiments at a given strain rate. Thus, for all deformation experiments at a given global strain rate, the average local strain rate in the post-yield regime is very nearly the same (within 10%).

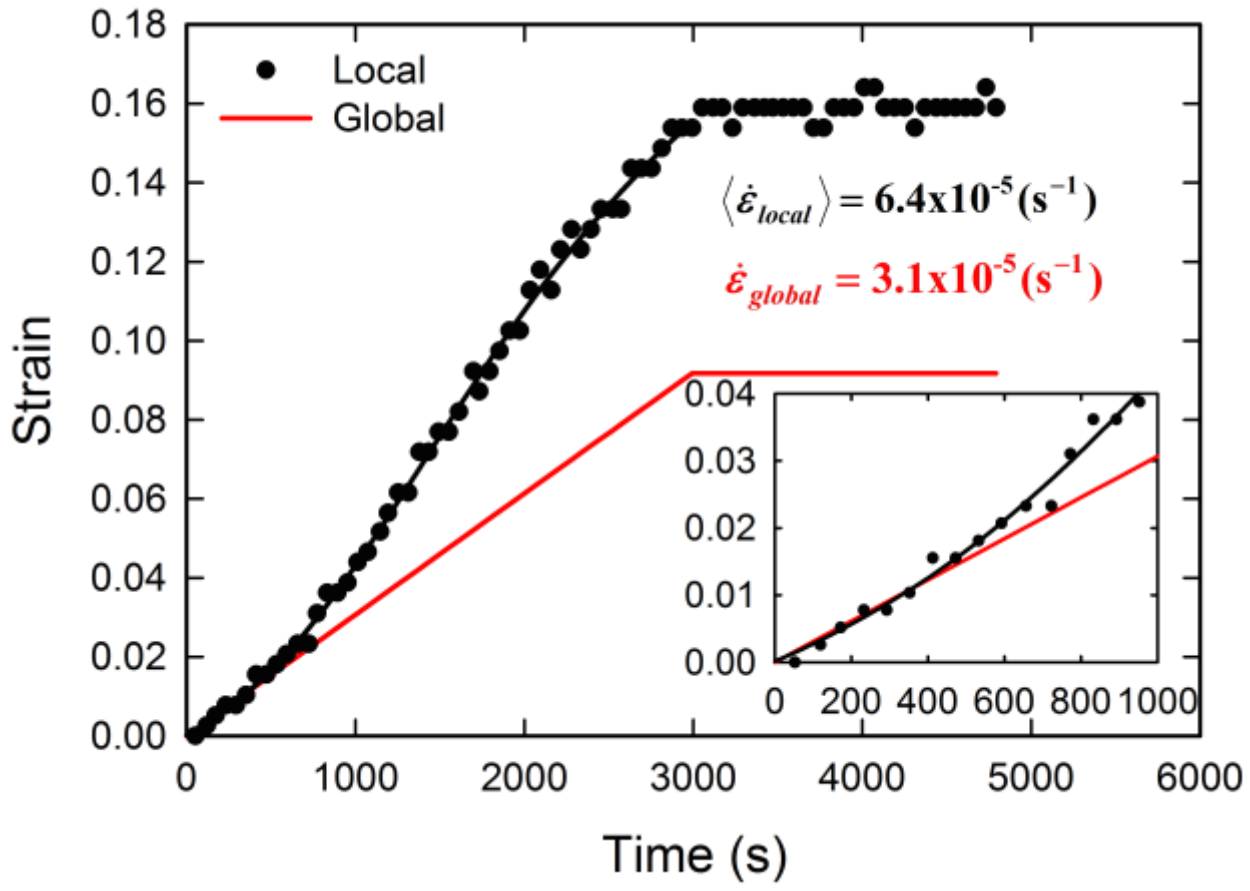

Figure 1. Evolution of local and global strain during deformation of a PMMA glass at 373 K. The local strain values (black points) are measured while the global strain (red line) is controlled. The inset shows that the local strain tracks the global strain closely until slightly prior to yield (which occurs at a strain of 0.028). After yield the local strain rate is about twice the global strain rate.

## Mechanical Relaxation Time

In this paper we discuss a mechanical relaxation time determined from a stress relaxation experiment performed after a constant strain rate deformation. While our procedure is directly based upon the work of Lee et al.,[27] many other investigators have used stress relaxation to infer segmental mobility.[25, 31] From the initial decay of the stress during relaxation, a mechanical relaxation time ($\tau_{mech}$) is defined as follows:

$$\frac{1}{\tau_{mech}} \equiv -\frac{1}{\sigma}\frac{d\sigma}{dt}\bigg|_{t=0} \qquad (1)$$

Here t=0 is the starting time of the stress relaxation experiment. Lee et al. proposed that this relaxation time is a measure of segmental mobility during constant strain rate deformation just prior to the initiation of the stress relaxation experiment.[27] Caruthers and coworkers used an alternate procedure to interpret stress relaxation experiments in which the stress was corrected by the equilibrium stress in the rubber state.[29] This alternate procedure yielded essentially identical mechanical relaxation times for our samples and will not be discussed further.

Equation 1 is the operational definition of one relaxation time that can be determined from a mechanical experiment. The relationship between $\tau_{mech}$ determined in this way and segmental mobility during nonlinear deformation is tested in this work. In the linear viscoelastic regime, $\tau_{mech}$ determined through eq. 1 will vary as a function of the strain at which stress relaxation is initiated, even though the average segmental relaxation time is not changing;[29] this dependence on strain will be discussed below. Related considerations in the determination of a mechanical relaxation time have been discussed previously.[36]

In our experiments, the error in a single measurement of $\tau_{mech}$ is estimated to be +/- 10% from the consistency with the overall trend of the measurement series. This error is about the size of the symbols for $\tau_{mech}$ in the figures below.

## Probe Reorientation Measurements of Segmental Mobility

We performed optical experiments that measure the reorientation of an ensemble of molecular probes (DPPC) dispersed in the PMMA films. This technique has been described previously[8, 33] and only a brief account is provided here. The key assumption of this approach is that probe reorientation is a good reporter of segmental dynamics in the glass during deformation. In the absence of deformation, reorientation of probes (on the time scale of seconds) has been previously shown to accurately monitor the segmental dynamics of polymer melts near $T_g$.[37-39]

The probe reorientation experiment begins with a polarized photobleaching beam that creates an oriented set of unbleached probes in a small portion of the sample. The polarized fluorescence of these molecules in response to a weak circularly polarized reading beam allows the time-dependence of probe reorientation to be determined.[8, 33] Quantitatively, we determine the anisotropy decay function *r(t)* associated with the second Legendre polynomial. Fitting the anisotropy decay with a stretched exponential function, we extract $\tau_{probe}$ as a measure of the segmental relaxation time in the sample.

$$r(t)/r(0) = e^{-(t/\tau_{probe})^{\beta}} \quad (2)$$

The exponent β characterizes the extent to which the anisotropy decays exponentially. In our previous work,[8, 11] we reported the integral of the correlation function, i.e., the rotational correlation time $\tau_c$. For the present experiments, we have found that the 1/e time ($\tau_{probe}$) is a more consistent reporter of probe reorientation as long as we observe at least 50% of the anisotropy decay. In order to meet this condition, the time used to record a single anisotropy decay was varied from 5 to 40 minutes.

## Results

### Stress Relaxation and Mechanical Relaxation Times

We performed constant strain rate deformations on lightly crosslinked PMMA glasses at three different strain rates. At each strain rate, stress relaxation was initiated after various strain levels were reached, in order to extract a mechanical relaxation time. Our procedure is illustrated in figure 2 for experiments with the global strain rate of $3.1 \times 10^{-5}$ $s^{-1}$. A high degree of reproducibility is seen in the deformation up to the strain at which the stress relaxation is started. To prevent systematic error, the experiments were not performed sequentially with respect to the strain value at the start of stress relaxation. The stress relaxation experiments were performed for at least 5400 seconds after the deformation stopped however the later portions were removed from the figure for clarity. Additional stress relaxation experiments were obtained as part of this sequence but are not shown in the figure, also for reasons of clarity.

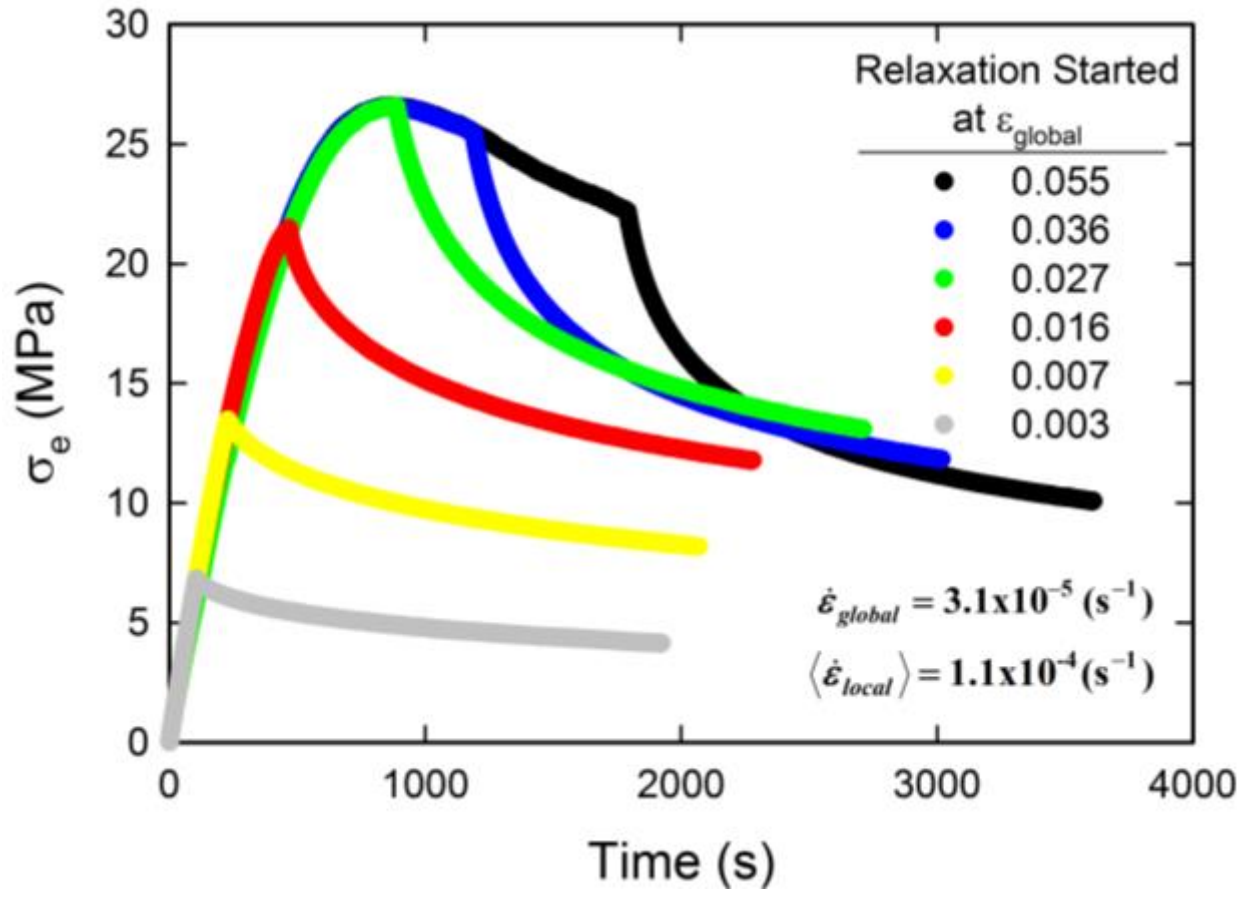

Figure 2. Engineering stress plotted against time for several constant strain rate experiments (all with global strain rate of $3.1x10^{-5}$ $s^{-1}$) followed by stress relaxation. Stress relaxation was initiated at the strain levels indicated. The temperature was 373 K ($T_g$ – 19 K).

The mechanical properties observed during the constant strain rate deformations shown in figure 2 are similar to those reported in literature for PMMA glasses. Yield is observed at 27 MPa which is in the reported range of 20-27 MPa. [40-41] A modulus of 1700 ± 100 MPa is observed which is in the range of 1500-2500 MPa as reported in literature.[42] Finally the yield strain of 0.028 ± 0.002 is consistent with our previous work.[11]

Stress relaxation experiments were performed after constant strain rate deformations at three global strain rates: $3.1x10^{-5}$, $3.1x10^{-6}$, and $1.5x10^{-6}$ $s^{-1}$. In figure 3, we compare the constant-strain rate deformations for these three strain rates. The values of the yield stress follow the expected logarithmic trend with the strain rate, in good agreement with the results reported in our previous work.[11] This range of strain rates was chosen to attempt to match the range used by Kim et al.[29] while considering the limitations of our apparatus.

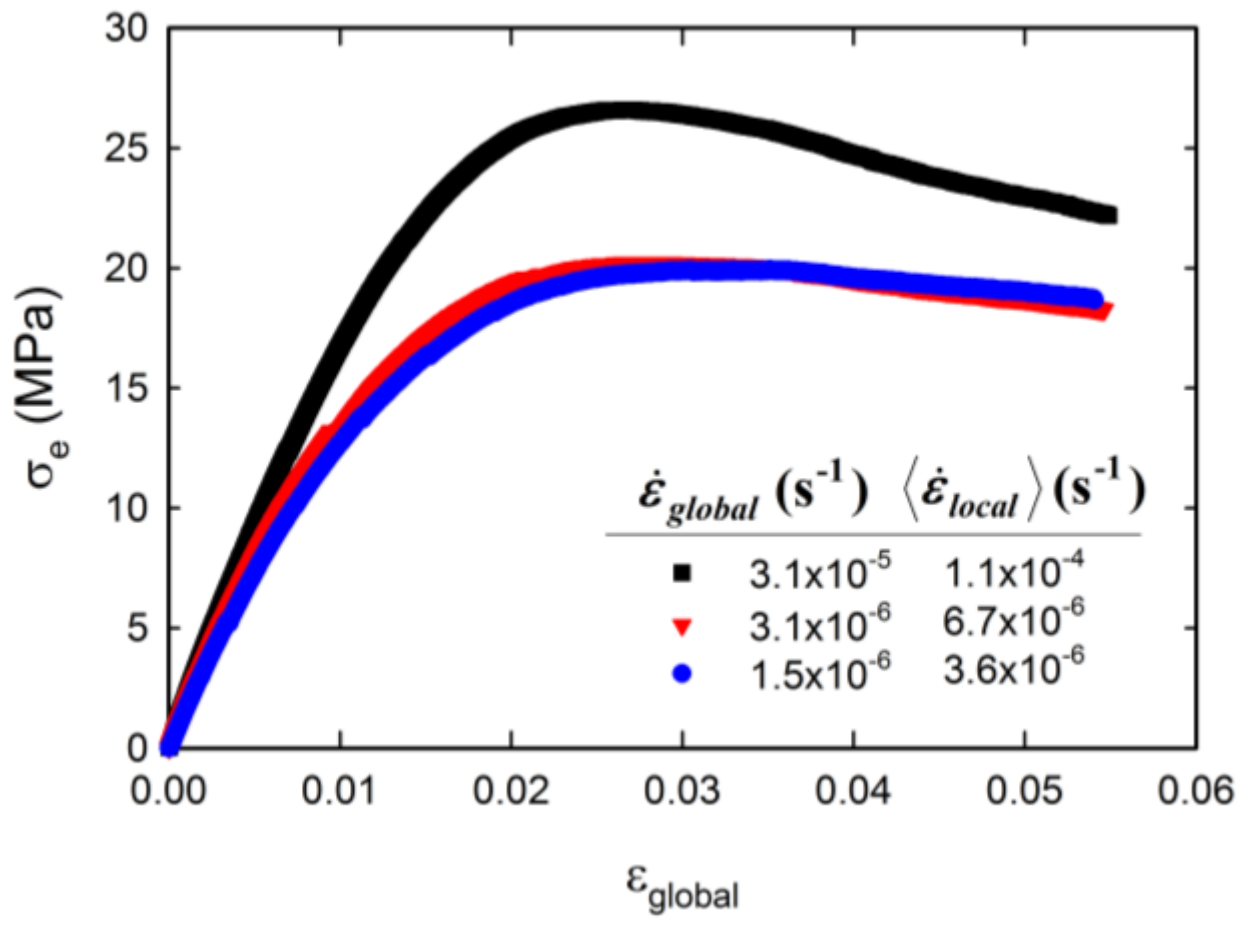

Figure 3. Stress as a function of global strain at the three strain rates used for stress relaxation experiments on PMMA glasses. Yield occurs at a global strain of 0.028 ± 0.002. The temperature was 373 K.

Figure 4 shows the stress relaxation curves for PMMA glasses starting at a wide range of strain values for two different global strain rates. In order to allow a direct comparison between the stress relaxation curves, we normalize the stress and shift the time to make the initiation of stress relaxation coincide. The relaxation curves at each strain rate show similar trends, with generally increasing decay rates as the strain increases until a fairly constant decay rate is exhibited slightly beyond the yield strain. In order to determine the mechanical relaxation time (see equation 1), we determine the slope of approximately the first 3% of the decay amount of the stress relaxation. For example, the slopes of the relaxation curves for 3.1 x10$^{-5}$ s$^{-1}$ strain rate were determined using the first 10 seconds of data. Some relaxation curves are not reported in figure 4 for clarity.

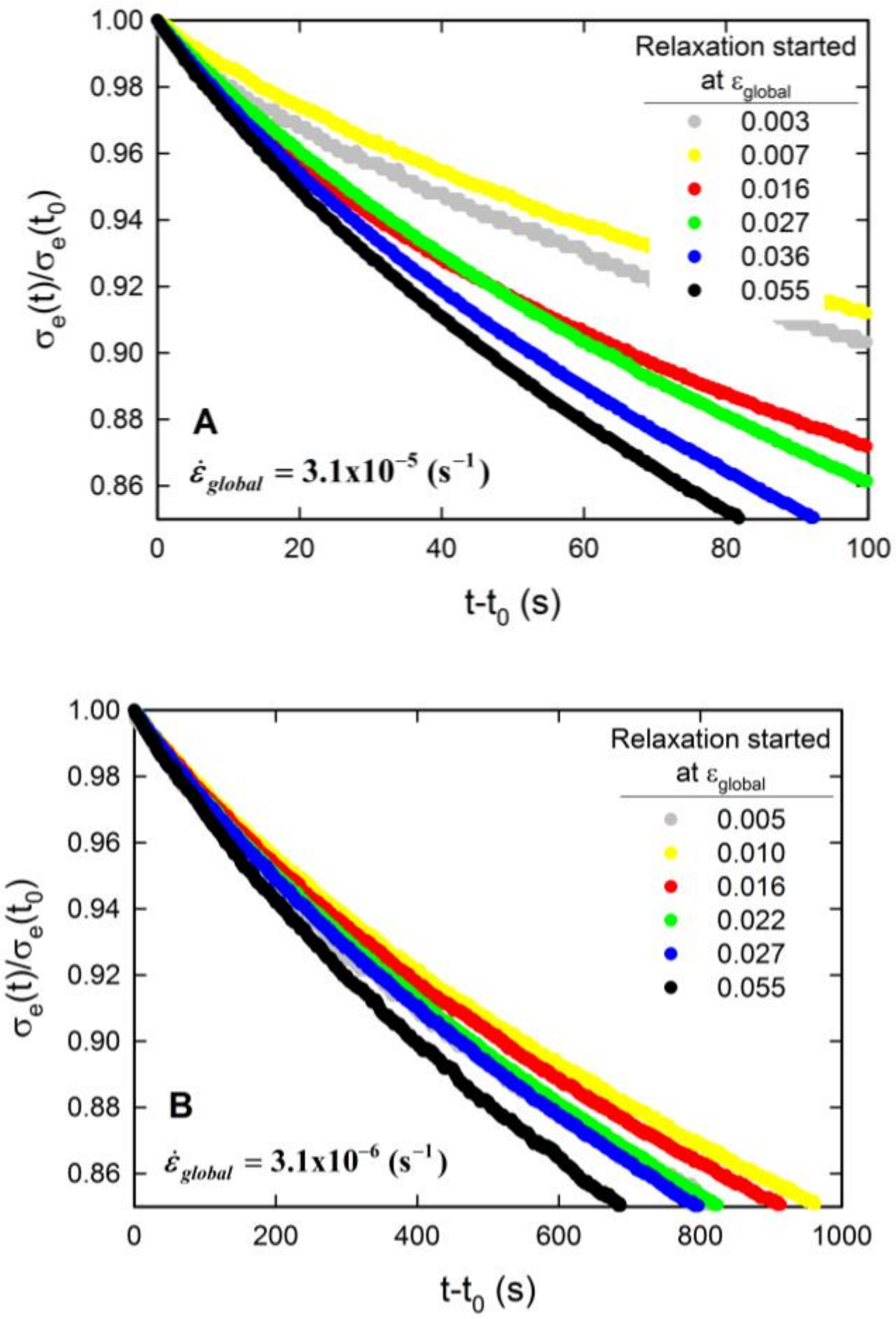

Figure 4. Normalized stress relaxation curves following constant strain rate deformation at A) $3.1x10^{-5}$ and B) $3.1x10^{-6}$ $s^{-1}$. Stress relaxation experiments started at the global strains indicated. Mechanical mobility is determined from the slope using the first 10 seconds of data for panel A and the first 100 seconds of data for panel B. The temperature was 373 K.

Mechanical relaxation times ($\tau_{mech}$) were determined from the stress relaxation curves using the definition provided in equation 1 and are reported in the lower portion of figure 5. $\tau_{mech}$ values are presented as a function of the global strain, normalized to the yield strain. The values of $\tau_{mech}$ decrease with increasing strain rate. For the two highest strain rates, $\tau_{mech}$ increases and then decreases with increasing strain. The lowest strain rate shows nearly constant mechanical relaxation times throughout the deformation. Lee et al. previously used equation 1 to obtain mechanical relaxation times for a PMMA glass deformed at similar strain rates ($1.2x10^{-5}$ to $7.5x10^{-4}$ s) and temperatures ($T_g$ – 20 K);[27] the results presented here show the same qualitative features.

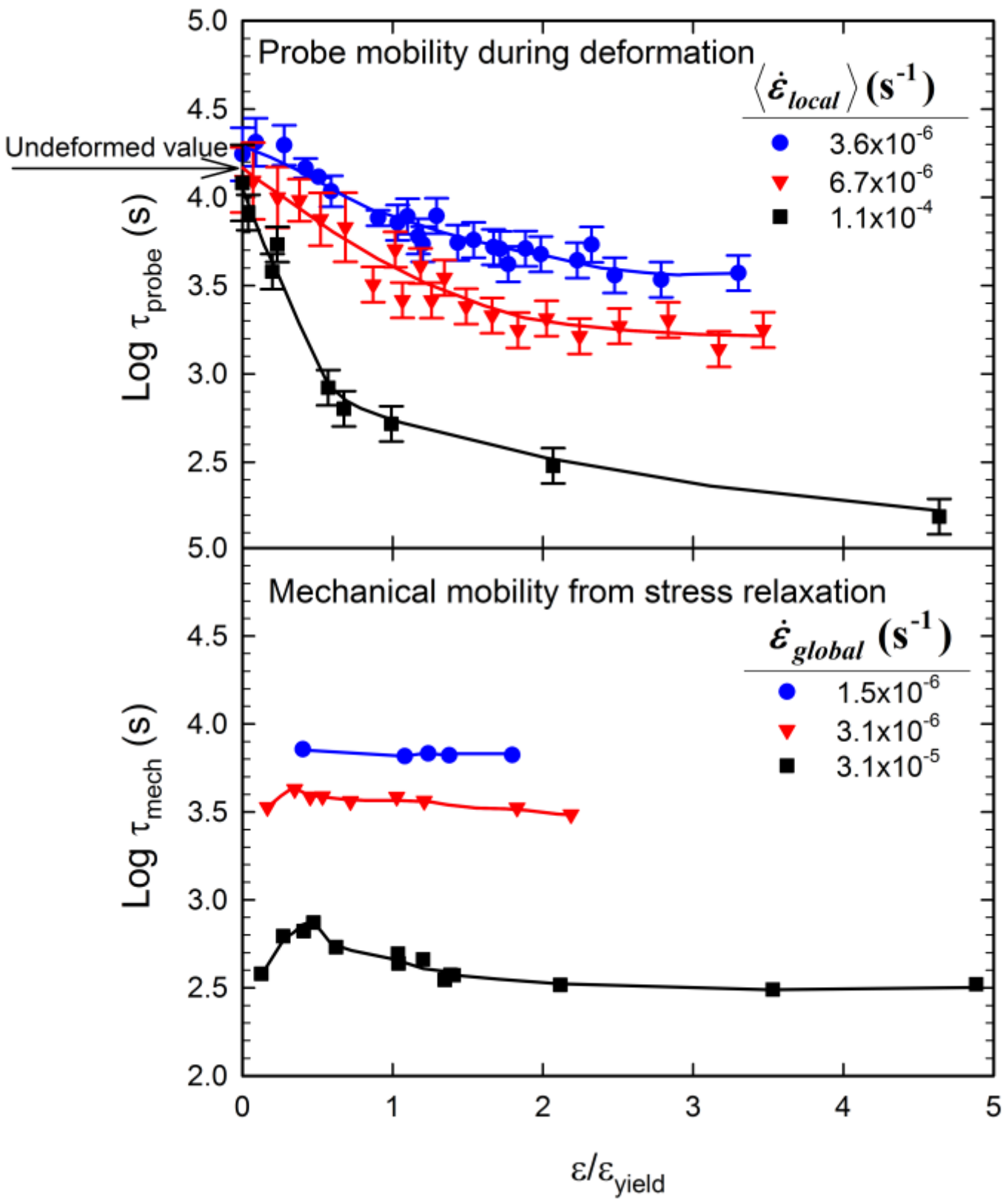

Figure 5. Probe and mechanical relaxation times plotted against normalized strain. The upper panel shows the probe relaxation time $\tau_{probe}$ measured during constant strain rate deformation. The lower panel shows the mechanical relaxation time as determined by the initial slope of the stress relaxation. Error bars on the probe relaxation times indicate the standard deviation of 3-8 independent experiments; error bars on mechanical relaxation times are estimated to be about the size of the symbols. Lines are provided as guides to the eye. The x-axis is local strain in the upper panel and global strain in the lower panel.

## Optical Measurements of Probe Reorientation

We performed optical measurements of probe reorientation during constant strain rate deformation and stress relaxation in order to characterize segmental mobility. In this method, an anisotropic distribution of probe molecules is created by a linearly polarized photobleaching beam and fluorescence is used to monitor their subsequent reorientation. The anisotropy decay function quantifies the reorientation of the ensemble of probes dispersed in the PMMA glass.

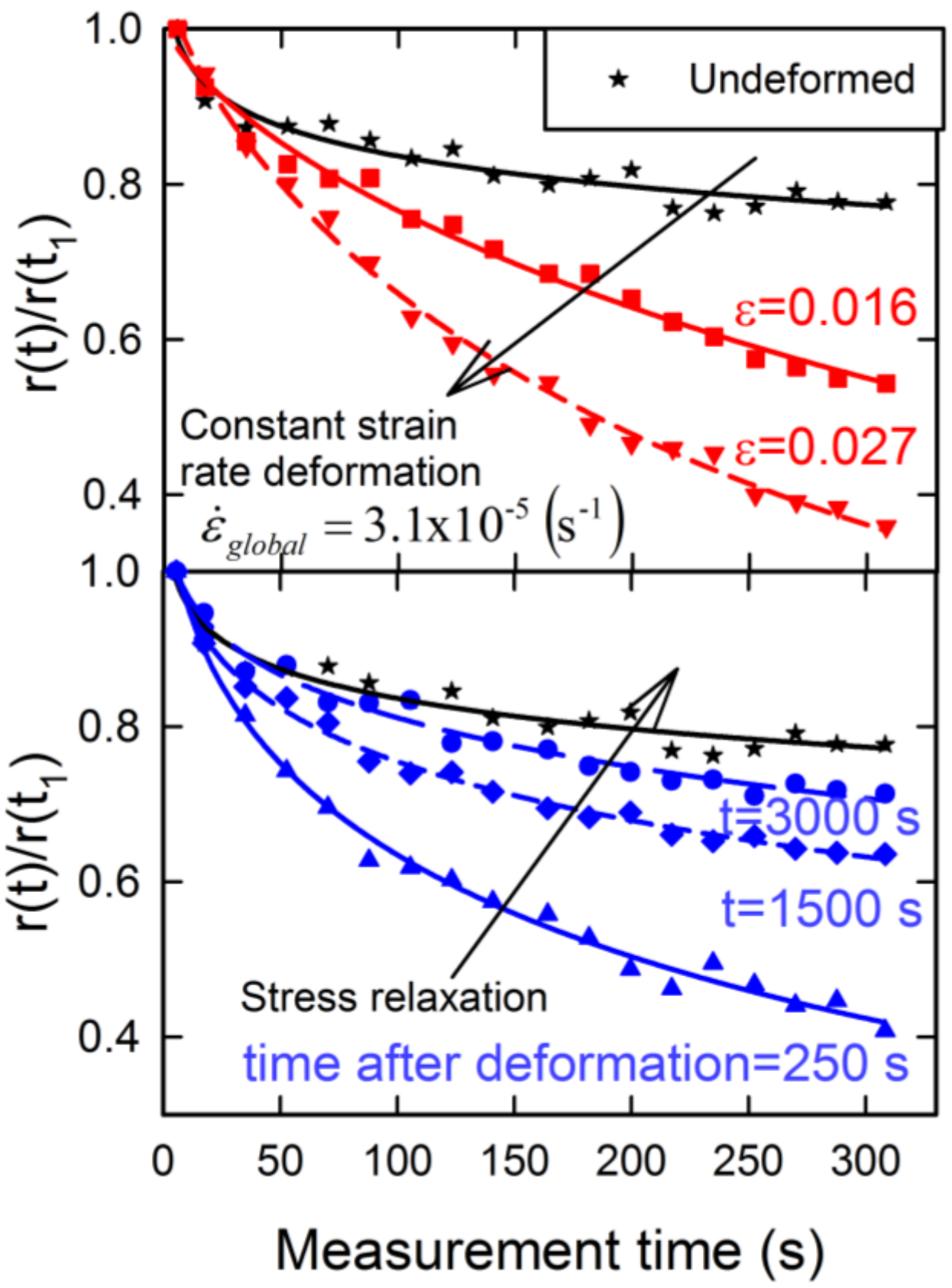

Figure 6. Anisotropy decay curves characterizing probe reorientation during constant strain rate deformation (upper panel) and during stress relaxation (lower panel). The anisotropy decays more rapidly as the strain increases during deformation, indicating increasing mobility. During stress relaxation, the anisotropy decays more slowly with increasing time, indicating that mobility is decreasing towards that of the undeformed glass. Solid lines are fits to the stretched exponential function. In this experiment, stress relaxation was initiated at a strain of 0.027.

In the upper panel of figure 6, representative anisotropy decay curves obtained during constant strain rate deformation are presented. The black stars indicate the anisotropy decay for the sample just prior to deformation. As the strain increases, the anisotropy decays more rapidly as indicated by the arrow, indicating that $\tau_{probe}$ becomes shorter and the ensemble of probe molecules reorients more rapidly. In the lower panel of figure 6, we show representative anisotropy curves that were measured during the stress relaxation portion of the experiment. As the relaxation proceeds (indicated by the arrow), the anisotropy curves decay more slowly, indicating that probe reorientation is slowing and that the mobility is returning to that of the undeformed sample. Each anisotropy curve is fit with the stretched exponential function (equation 2) to determine the probe relaxation time ($\tau_{probe}$) and the stretched exponential parameter ($\beta$).

The top portion of figure 5 shows probe relaxation times plotted against the normalized strain; these relaxation times were obtained during the constant strain rate portion of the experiment. For all three strain rates, the probe relaxation time starts very near the relaxation time observed for the undeformed sample [$\log(\tau_{probe}/s)=4.2$]. As the strain increases, the probe relaxation time shortens as yield is approached and then changes more slowly in the post-yield regime. The decreases in the probe relaxation time are more pronounced at higher strain rate.

The results presented in the top portion of figure 5 are similar to those reported in a previous study but extend the range of strain rates explored by about an order of magnitude to smaller values.[11] This previous work showed that small changes of the probe relaxation time in the post-yield regime were due to small variations in the local strain rate and that correction for this effect yielded probe relaxation times that were constant in the post-yield regime. The major features of the results are consistent with theory[13] and simulations[21]. At the lowest strain rate, the evolution of $\tau_{probe}$ is not monotonic with strain; this is a reproducible feature of the data (see lowest strain rate data in Figure 2a of reference[43]) but we are not confident that we understand its physical origin.

We also measured the probe relaxation time during the stress relaxation portion of each experiment as illustrated in the bottom portion of figure 6. In figure 7, we show the evolution of the probe relaxation time during constant strain rate deformation at $3.1 \times 10^{-5}$ $s^{-1}$ and then during three of the stress relaxation experiments. The data acquired during constant strain rate deformation (black points) were previously shown in the top portion of figure 5. The new information in figure 7, acquired during stress relaxation, is shown by the red, green, and blue points; for each stress relaxation experiment, the vertical colored lines indicate the time ($t_0$) at which stress relaxation was initiated. For example, for the stress relaxation experiment at a local strain value of 0.027 (green curve), the experiment begins with deformation decreasing the probe relaxation time from the undeformed value. At $t_0$ = 870 s, the deformation was stopped and stress relaxation began. The first probe measurement during stress relaxation indicates a slowdown of molecular motion relative to that during the constant strain rate portion of the experiment; this is indicated by the green arrow. As stress relaxation proceeds, the probe relaxation time continues to increase towards the undeformed value as expected for a sample undergoing physical aging. Similar qualitative trends are observed for stress relaxation experiments initiated at the other two strain levels.

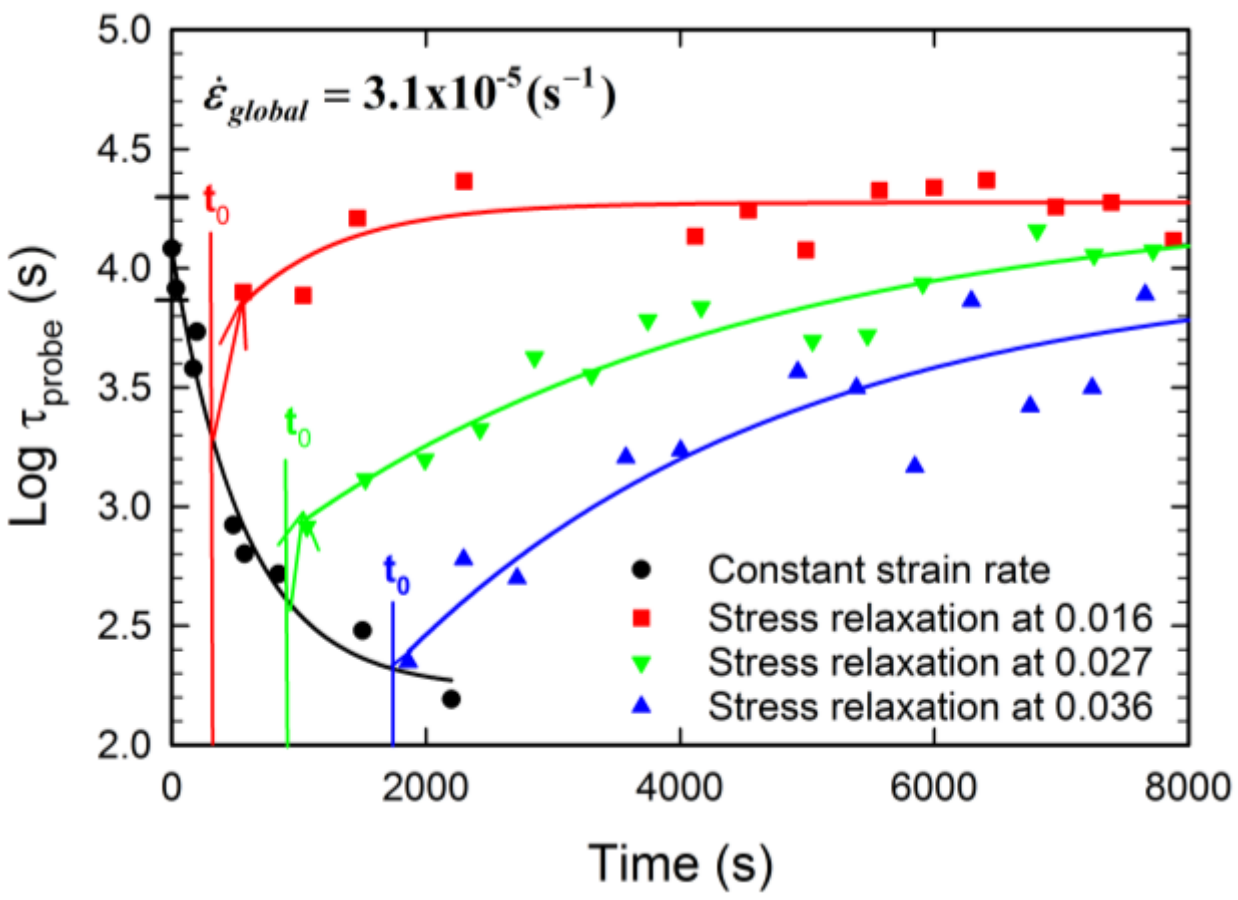

Figure 7. Probe relaxation times measured during constant strain rate deformation (black points) and during stress relaxation (colored points) initiated at the strains and times ($t_0$) indicated. During stress relaxation, the relaxation time increases toward the relaxation time of the undeformed glass, i.e., segmental mobility decreases. Lines are provided as guides to the eye.

Dynamical Heterogeneity during Constant Strain Rate Deformation

During the probe mobility measurements, we are also able to obtain a measure of the distribution of the segmental relaxation times from the non-exponentiality of the probe anisotropy decay function, as quantified by the KWW β parameter. The observed behavior is shown in figure 8 for the three strain rates utilized here. At the highest strain rate, there is a strong tendency for β to increase from the value in the undeformed glass. The same trend is observed to a lesser extent at the lower strain rates.

We interpret figure 8 to indicate that segmental dynamics become more homogenous during deformation in the pre-yield regime. Larger β values correspond to a narrower distribution of relaxation times. Increasing the strain rate makes the narrowing of the distribution of relaxation times more pronounced. This is the first study in which the evolution of the distribution of relaxation times during constant strain rate deformation can be clearly observed.

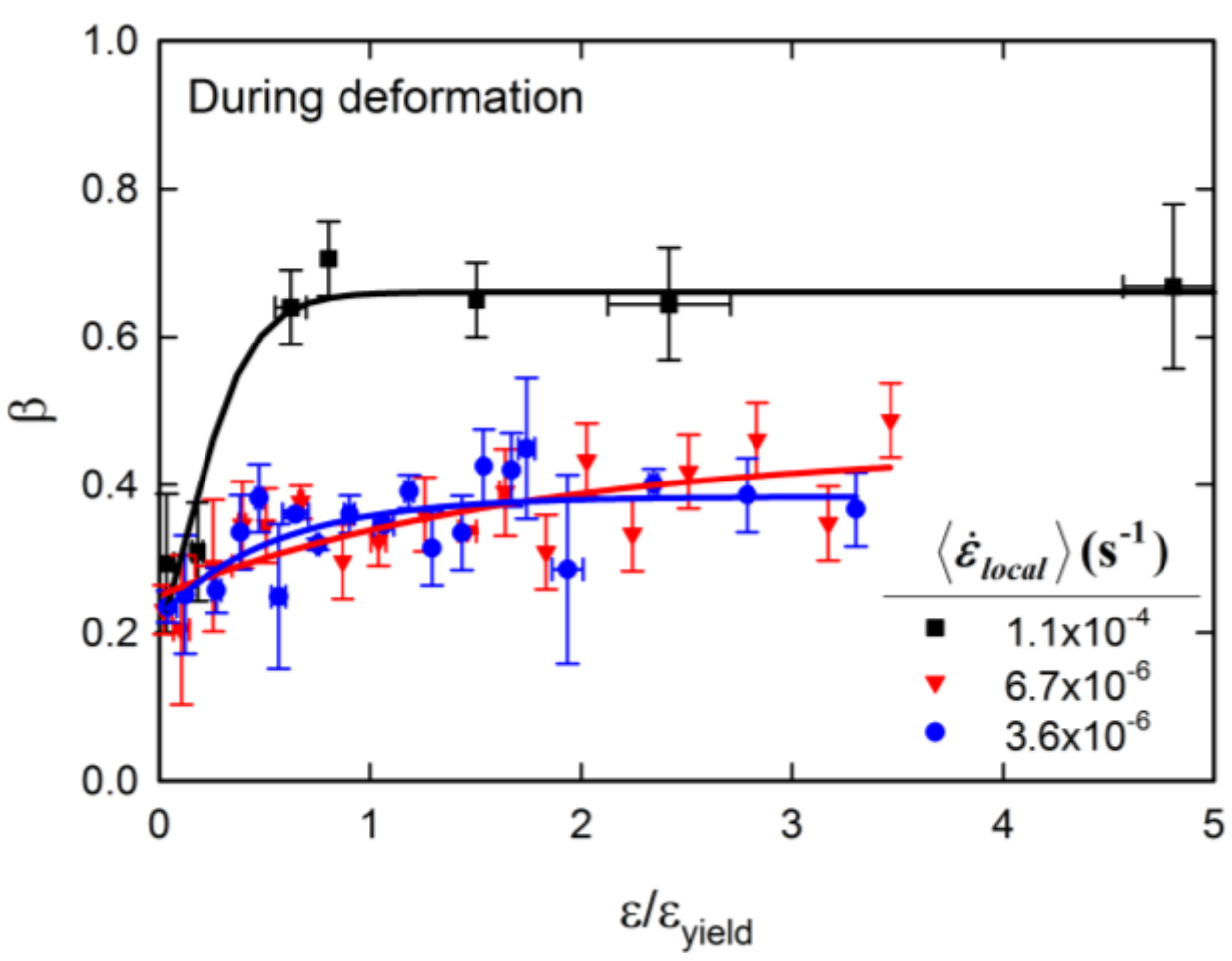

Figure 8. The KWW β parameter observed for probe reorientation during constant strain rate deformation as a function of the normalized strain. As deformation proceeds, β increases indicating a narrowing the distribution of relaxation times, consistent with the view that the sample becomes more dynamically homogenous. Error bars indicated are the standard deviation of 3-6 measurements. Lines are provided as a guide to the eye.

**Discussion**

In this section, we compare the probe relaxation times determined by the optical method with mechanical relaxation times determined through the definition shown in equation 1, separating our discussion of the pre-yield and post-yield regimes. We also consider the evolution of dynamical heterogeneity during constant strain rate deformation.

Mechanical and Probe Relaxation Times: Pre-yield Regime.

Figure 5 compares the probe and mechanical relaxation times at three different strain rates. Clearly the two observables evolve in different ways as the strain increases. Figure 9 allows a more detailed comparison for one strain rate. The most striking differences between the mechanical and probe relaxation times are observed in the pre-yield regime. The probe relaxation time changes by 1.5 decades as yield is approached while the mechanical relaxation time changes less than 0.3 decades. The different evolution of the mechanical and probe relaxation times prior to yield cannot be explained by differences in the local and global strain rates, as figure 1 shows that these are nearly identical in the pre-yield regime.

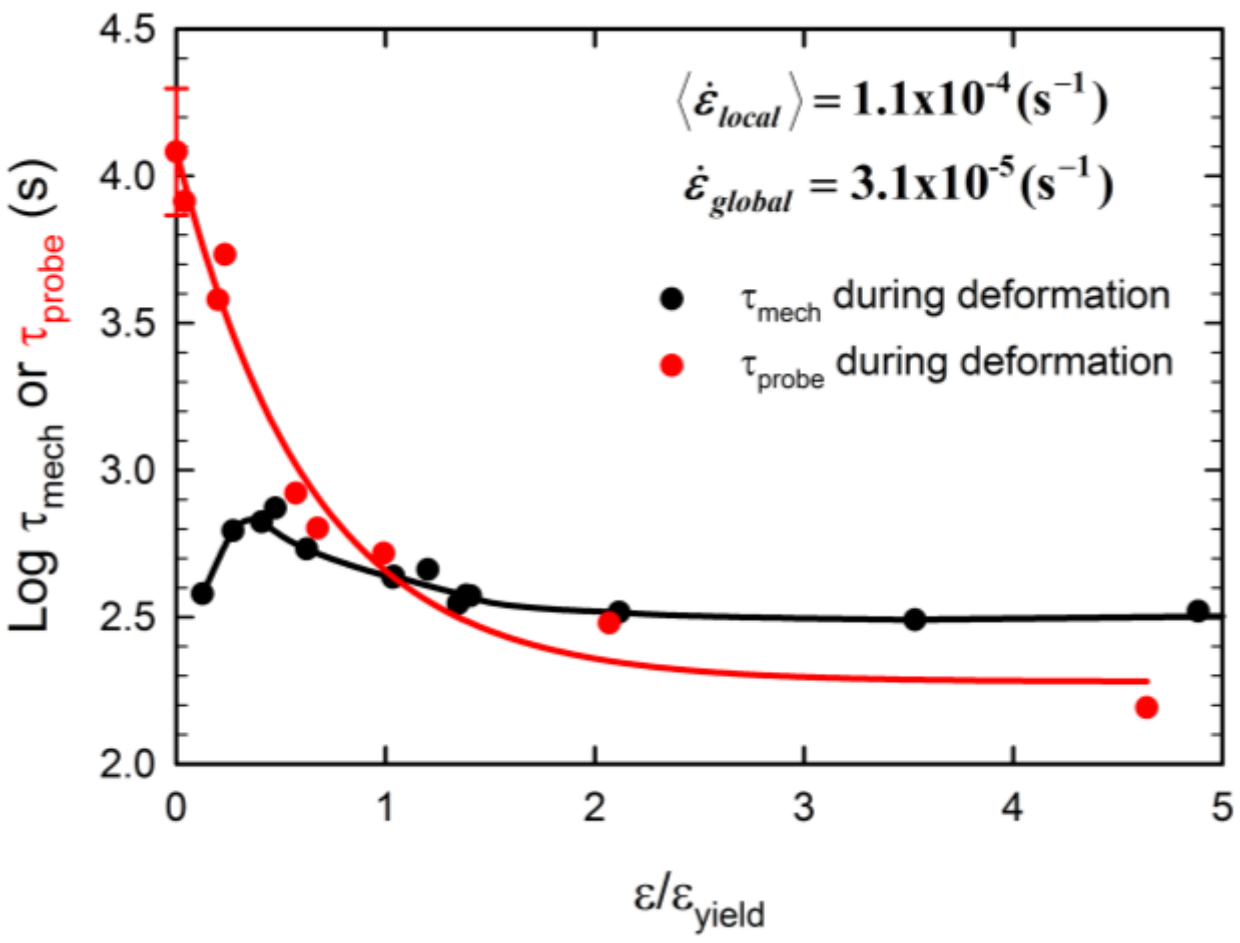

Figure 9. Comparison of the mechanical relaxation time (black) and the probe relaxation time (red) during a constant strain rate deformation. The probe relaxation time

decreases by a factor of 130 whereas the mechanical relaxation time only changes by a factor of 3. In the post-yield region. The mechanical and probe relaxation times exhibit similar trends. Error bars for the optical and mechanical relaxation times are estimated as +/- 0.15 decades and +/- 0.05 decades, respectively. The x-axis is local strain for $\tau_{probe}$ and global strain for $\tau_{mech}$.

Figure 5 shows that, at small deformations, the probe relaxation times for all strain rates are very similar to the relaxation time of the undeformed sample. This is expected at the very beginning of the constant strain rate experiment since this will be a linear response regime and in this limit the molecular relaxation times will not be altered by deformation. A similar argument must apply to the mechanical relaxation times but apparently the smallest strains investigated (0.003) are too large to be in the linear response regime. Viewed from this perspective, the mechanical relaxation times must also show a large change in the pre-yield regime that unfortunately was not observed in our experiments. The gradual shortening of the probe relaxation time observed here as yield is approached is qualitatively consistent with the behavior of the segmental relaxation time in computer simulations of polymer glasses[21] and with the molecular theory of Chen and Schweizer.[13] In contrast, the mechanical relaxation time approximately reaches its post-yield value considerably before yield.

We considered the possibility that the mechanical relaxation time (which is measured after deformation is stopped) might better correlate with the probe relaxation time observed immediately after the initiation of stress relaxation. In figure 7, one can see that the first measurement of the probe relaxation time after the start of stress relaxation also changes by roughly 1.5 decades as yield is approached. Therefore this does not provide a way to reconcile the qualitative differences between the probe and mechanical relaxation times. We also considered that there might be a very fast mechanical relaxation process immediately following the switch from constant strain-rate deformation to stress relaxation. While figure 2 shows no evidence of such a fast mechanical relaxation, we considered this possibility more carefully by examining the stress measurements which are acquired every 200 ms. The first stress measurement during stress relaxation matches the last measurement during constant strain rate deformation to within 0.1%, meaning that any unobserved fast mechanical process would have to have an extremely small amplitude.

Mechanical and Probe Relaxation Times: Post-yield Regime.

As shown in figures 5 and 9, the probe and mechanical relaxation times behave in a qualitatively similar manner in the post-yield regime, with each relaxation time being approximately constant. In figure 10, we show a comparison of these two quantities. The direct comparison is complicated by the difference between the local and global strain rates; as discussed above and as shown in figure 1, the local strain rate is 2-3 times larger than the global strain rate after yield. In both panels of figure 10, the probe relaxation times are plotted against the local strain rate; this is the correct choice as the local mobility is influenced only by the local deformation characteristics. The probe relaxation times observed in these experiments are in good agreement with recently reported results[11] (shown as open symbols in figure 10). The top and bottom panels of figure 10 differ only in the presentation of the mechanical relaxation times. In the top panel, these are plotted against the global strain rate; in the bottom panel, these are plotted against the local strain rate. Deciding which x-axis is most appropriate depends upon the detailed nature of the inhomogeneous deformation that occurs in the post-yield regime; it would be useful to revisit this relationship for experiments in compression for which a more homogeneous deformation would be anticipated. Fortunately, the interpretation of the two panels in figure 10 is essentially the same; in both cases, there is a strong correlation between mechanical and probe relaxation times in the post-yield regime with at most a factor of ~2 difference between the absolute relaxation times at any given strain rate.

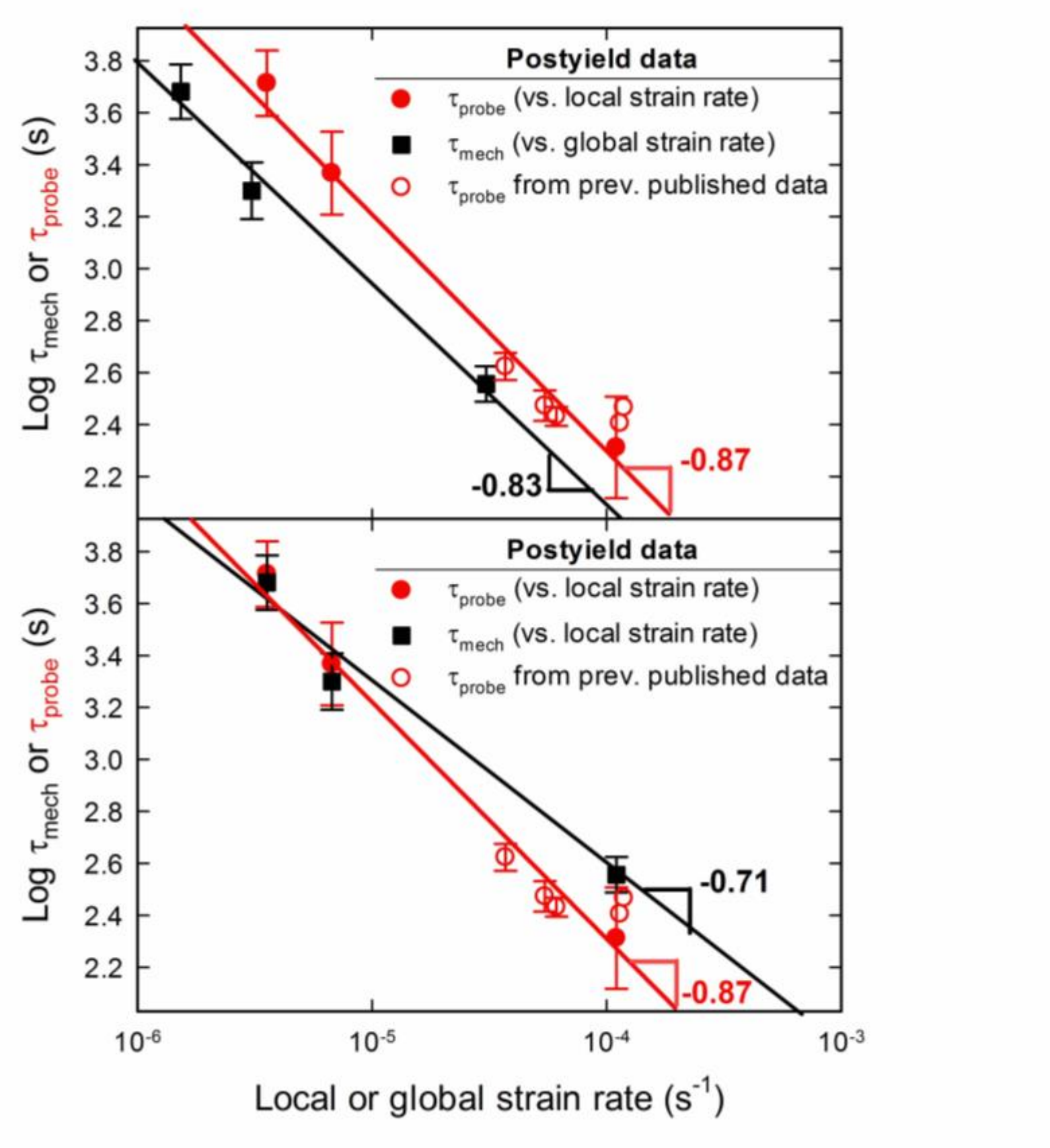

Figure 10. Mechanical and probe relaxation times from the post-yield regime, plotted against strain rate. In both panels, probe relaxation times are plotted against local strain rate. Mechanical relaxation times are plotted against the global and local strain rates in the upper and lower panels, respectively. In the post-yield regime, the two types of relaxation times show a similar dependence on strain rate. Previously reported constant strain rate data for probe reorientation is also shown (open symbols, from reference [13]). Error bars represent one standard deviation of the averaged post yield data.

In figure 10, both the probe and mechanical relaxation times exhibit a power law relationship with the strain rate, with the exponents of -0.8 +/- 0.1. This indicates that an order of magnitude increase in the strain rate causes almost an order of magnitude decrease in the relaxation times. Numerical solutions to the theory of Chen and Schweizer indicate an exponent of -0.86, in reasonable agreement with the observed behavior for both relaxation times.[13] Within the context of the theory, the smaller relaxation time at higher strain rates results partially from the increased stress required to maintain the higher strain rate (via an Eyring-like mechanism). The higher strain rate also changes the glass structure (as quantified by an increase in the amplitude of density fluctuations). Both of these mechanisms lower the barriers for segmental rearrangements. The experimental results for probe relaxation in figure 10 are also roughly compatible with an exponent of -1, as anticipated by the theory of Fielding, Larson, and Cates[15-16] and the mechanism envisioned by Chung and Lacks.[24]

The parallel behavior shown by the probe and mechanical relaxation times in figure 10 (particularly in the upper panel) is encouraging and suggests that both measurements are sensitive to the average segmental relaxation time in the post-yield regime. The offset between the two curves (at most a factor of two) can be tentatively explained as follows. The probe relaxation time observed in an equilibrium polymer melt above $T_g$ depends upon the size of the probe molecule, with probes over a range of sizes all exhibiting the strong temperature dependence of the segmental relaxation process.[39] We speculate that, in the equilibrium melt, the average relaxation time of the DPPC probe may be up to twice as long as the average segmental relaxation time of the neat polymer as detected in a linear mechanical experiment. This factor of two would then carry over into the non-equilibrium glass during deformation as shown in figure 10. This explanation could be tested by comparing probe and mechanical relaxation times during glass deformation experiments in the linear response regime.

The results in figure 10 are roughly consistent with very recent work by Liu et al., who performed stress relaxation experiments after constant strain rate deformation in glasses of polycarbonate, polystyrene, and poly(methyl methacrylate).[31] These authors reported that the stress relaxation curves exhibited a nearly universal decay after multiplying the observation time by the deformation rate used prior to stress relaxation (this result would indicate a slope of -1 for the black line in figure 10). Liu et al. interpreted their results in terms of enhanced segmental dynamics and this is fully consistent with our observations in the post-yield regime. Interestingly, Liu et al. also report successful scaling of their stress relaxation curves obtained prior to yield. Given the disconnect between probe reorientation and stress relaxation prior to yield (figure 9), it is not clear to us why this procedure should be successful. In their pre-yield experiments in which the strain rate was varied, Liu et al. choose the stress at the beginning of stress relaxation to be constant and this is likely an important feature.

Understanding the Differences Between Probe and Mechanical Relaxation Times.

We take the view that the probe relaxation times are proportional to the average segmental relaxation time during deformation. Supporting this view is the linear relationship between probe reorientation times and dielectric relaxation times for polymer melts in the absence of deformation[37-39], and the qualitative agreement of the probe relaxation times with simulation results[21-22] and the prediction of the theory of Chen and Schweizer.[13] Furthermore, probe relaxation times in the pre-yield regime have been previously obtained during creep deformation and found to be in good qualitative agreement with the Eyring prediction for stress-activated dynamics.[9] On the other hand, the identification of $\tau_{mech}$ (as defined in equation 1) as the timescale for segmental dynamics during deformation[27] is an assumption. Given the results from figures 5 and 9, if probe reorientation is accurately tracking the average segmental relaxation time during deformation, then this cannot be the case for $\tau_{mech}$.

For a material that is described by the *linear* viscoelastic model equations with a distribution of relaxation times, $\tau_{mech}$ (as defined by equation 1) increases monotonically with strain. This point is considered in detail in reference [29] and here we recapitulate some results established there. Consider a Maxwell model in which the individual elements are characterized by moduli $E_i$ and relaxation times $\tau_i$. Deformation at a constant strain rate $\dot{\epsilon}$ until a time $t_1$ is followed by stress relaxation. The normalized stress relaxation response as a function of time $\tilde{t}$ (= $t - t_1$) is equal to[29]:

$$\bar{\sigma} = [\sum_{j=1}^{n} E_j \tau_j \left(1 - e^{-\frac{t_1}{\tau_j}}\right)]^{-1} \ [\sum_{i=1}^{n} E_i \tau_i e^{-\frac{\tilde{t}}{\tau_i}} \left(1 - e^{-\frac{t_1}{\tau_i}}\right)] \qquad (3)$$

For our purposes, the most important term in this equation is the final term in parentheses in the numerator. Note that as the time of the constant strain rate deformation increases (i.e, increasing $t_1$), this term goes to zero for short relaxation times but not for long relaxation times. Using equation 3, one can show that if stress relaxation is initiated at very large strains, $\tau_{mech}$ will be equal to the average relaxation time $<\tau> = [\sum E_i \tau_i]/\sum E_i$, where $E_i$ and $\tau_i$ specify the modulus and relaxation time of an individual Maxwell element. On the other hand, if stress relaxation is initiated after an infinitesimally small strain, $\tau_{mech}$ will be equal to $<1/\tau>^{-1} = \sum E_i/[\sum E_i / \tau_i]$. Note that these results, obtained for *linear* viscoelasticity, predict that $\tau_{mech}$ will increase considerably with strain; for a relaxation time distribution with a breadth of three decades, the increase in $\tau_{mech}$ with strain will be more than two decades. Unfortunately, there are no similarly rigorous results for nonlinear deformation.

We speculate that the behavior observed for $\tau_{mech}$ in figure 9 is the result of two competing effects that influence $\tau_{mech}$ in opposite directions. In our nonlinear deformation experiments, the segmental relaxation time (as monitored by the probe) decreases by roughly two decades as the strain increases towards yield, and we expect this will act to decrease $\tau_{mech}$. On the other hand, we expect an increase in $\tau_{mech}$ with strain due to the effect described in the previous paragraph. Combining these effects, one expects two competing trends in the mechanical relaxation time as a function of strain: We speculate that these two influences largely compensate each other in our experiments, leading to the observed behavior of $\tau_{mech}$ being almost independent of strain during non-linear deformation. On the other hand, the probe relaxation time is only sensitive to the non-linear effect, and thus $\tau_{probe}$ and $\tau_{mech}$ have a different dependence upon strain.

Support for this speculative interpretation of $\tau_{mech}$ is found in figure 10. According the idea of the previous paragraph, $\tau_{mech}$ at large strains should approximately equal the average segmental relaxation time even during a nonlinear deformation that has strongly perturbed the original relaxation time distribution. The near equality of $\tau_{mech}$ and $\tau_{probe}$ shown in figure 10 is consistent with this. On the other hand, the non-monotonic behavior of $\tau_{mech}$ shown in figure 5 does not seem to follow simply from this view. We leave it as a challenge for theory and simulations to provide a better explanation of the behavior of $\tau_{mech}$.

While our experiments indicate that $\tau_{mech}$ is not an accurate indication of the average segmental relaxation time during nonlinear deformation, we view $\tau_{mech}$ as an important characterization of the state of the system during deformation. The fact that $\tau_{mech}$ and $\tau_{probe}$ are qualitatively different in the pre-yield regime indicates that these combined observables are likely a stringent test for models of polymer glass deformation. The stochastic constitutive model of Medvedev and Caruthers, which utilizes a distribution of relaxation times, should be able to provide predictions for both $\tau_{mech}$ and $\tau_{probe}$, as well as the changes in the breadth of the relaxation time distribution that occur during deformation. [14, 30]

Dynamical Heterogeneity During Deformation

Over the last two decades, a great deal of evidence has been amassed indicating that segmental dynamics in polymer melts (and more generally the $\alpha$ relaxation process in glassformers) is spatially heterogeneous near $T_g$.[44-45] In light of these results, we interpret the KWW $\beta$ parameter for our PMMA glass in the absence of deformation as a rough indicator of the extent to which dynamics are spatially heterogeneous.  From this perspective, the increase in the $\beta$ parameter during deformation can only mean the dynamics during deformation exhibits less spatial heterogeneity. While earlier work with the probe reorientation method indicated that the distribution of segmental relaxation time narrows as a result of deformation, figure 8 is the first data showing the evolution of this effect during constant strain rate deformation.

The possibility that the distribution of relaxation times might narrow during deformation is not considered in many theories and models,[5, 13] but this result has been observed in simulations[46] and experiments on colloids.[47] Recently Medvedev and Caruthers[14, 30] developed a stochastic constitutive model to describe the deformation of polymer glasses.  A unique feature of this model is the incorporation of a distribution of relaxation times that stems from the spatially heterogeneous dynamics of real polymer glasses. This model has been used to predict the mechanical response of PMMA glass to uniaxial compression and extension.  This model predicts that constant strain rate deformation leads to a decrease in the average segmental relaxation time.  In addition, the model predicts that the distribution of segmental relaxation times narrows in the pre-yield regime and becomes constant in the post-yield regime. The experimental results presented here are in qualitative agreement with these three predictions.

## Conclusion

In this work we compared two relaxation times obtained during the constant strain rate deformation of a polymer glass. A purely mechanical relaxation time obtained from the initial rate of stress relaxation has been compared with the reorientation time for a molecular probe. Both relaxation times decreased significantly as a result of deformation with the probe relaxation time changing by about a factor of 100 at the highest strain rate. While there is a good correlation between the probe and mechanical relaxation times in the post-yield regime, these two relaxation times behave qualitatively differently prior to yield. The probe relaxation time decreases smoothly from its undeformed value as yield is approached while the mechanical relaxation time is nearly constant at its post-yield value. Based upon comparison with computer simulations and theory, we view the probe relaxation times as accurately representing the average segmental relaxation time of the polymer during deformation. We speculate that the mechanical relaxation times change during deformation both as a result of the broad relaxation time distribution and due to enhanced segmental mobility. We anticipate that the combination of probe and mechanical relaxation times, together with the observed changes in the width of the distribution of relaxation times, will be an important test for models and theories of polymer glass deformation. Finding a purely mechanical experiment that is only sensitive to the changes in segmental mobility during deformation remains an important challenge.

## Acknowledgements

We thank the National Science Foundation (DMR-1104770 and DMR-1404614) for the support of this research. We thank Ken Schweizer, Kelly Hebert, Josh Ricci, James Caruthers, and Grigori Medvedev for helpful discussions. We thank Lian Yu and Travis Powell for assistance with DSC measurements.

# Comparison of mechanical and molecular measures of mobility during constant strain rate deformation of a PMMA glass

Benjamin Bending and M.D. Ediger*

## For Table of Contents Use Only

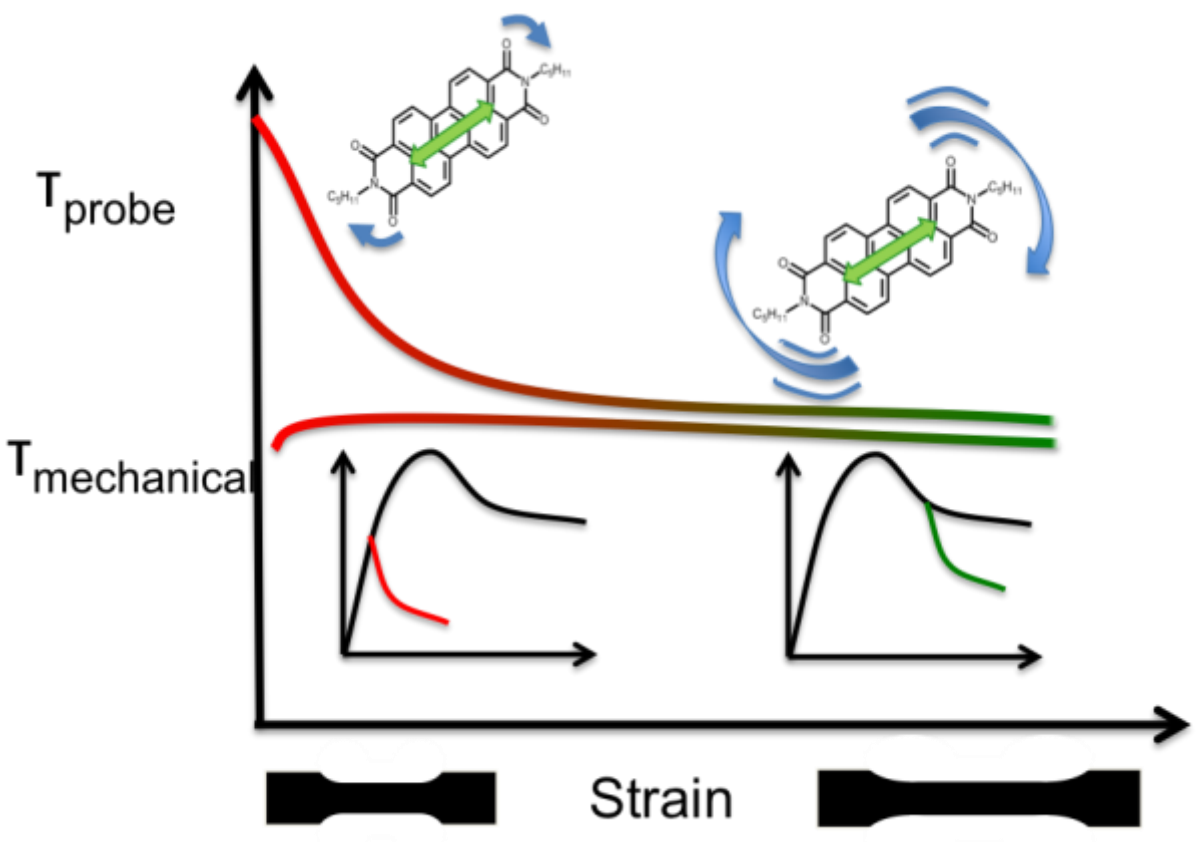

**Text for TOC figure:**

**When polymer glasses are subjected to nonlinear deformation, flow occurs as a result of enhanced segmental mobility. Understanding and predicting segmental dynamics during deformation is a key challenge. Here the authors test whether a purely mechanical relaxation time (obtained from stress relaxation) tracks changes in segmental dynamics during constant strain rate deformation. The mechanical relaxation time is found to be a reasonable approximation for the segmental relaxation time in the post-yield regime but fails significantly prior to yield.**